\documentclass[aps,showpacs,amsfonts,nofootinbib,preprintnumbers,nobalancelastpage]{revtex4}
\usepackage{graphicx}
\usepackage{epsfig}
\usepackage{xcolor}
\usepackage{rotating}
\usepackage{amsmath}
\usepackage{braket}
\usepackage{pdflscape}

\newcommand{\be}{\begin{equation}}
\newcommand{\ee}{\end{equation}}
\newcommand{\tr}{{\rm Tr}}

\def\cF{{\cal F}}

\def\cP{{\cal P}}

\begin{document}
\hfill{YITP-SB-15-34}

\title{Inverse Amplitude Method for Perturbative Electroweak
Symmetry Breaking Sector: The Singlet Higgs Portal as a Study Case}
\author{Tyler Corbett}
\email{corbett.t.s@gmail.com}
\affiliation{%
  C.N.Yang Institute for Theoretical Physics, SUNY at Stony Brook,
  Stony Brook, NY 11794-3840, USA}
\affiliation{ARC Centre of Excellence for Particle Physics at the Terascale,
School of Physics, The University of Melbourne, Victoria 3010, Australia}
\author{O.\ J.\ P.\ \'Eboli}
\email{eboli@fma.if.usp.br}
\affiliation{Instituto de F\'{\i}sica,
             Universidade de S\~ao Paulo, S\~ao Paulo -- SP, Brazil.}
\author{M.\ C.\ Gonzalez--Garcia} \email{concha@insti.physics.sunysb.edu}
\affiliation{%
  Instituci\'o Catalana de Recerca i Estudis Avan\c{c}ats (ICREA),}
\affiliation {Departament d'Estructura i Constituents de la Mat\`eria, 
Universitat
  de Barcelona, 647 Diagonal, E-08028 Barcelona, Spain}
\affiliation{%
  C.N.~Yang Institute for Theoretical Physics, SUNY at Stony Brook,
  Stony Brook, NY 11794-3840, USA}

\begin{abstract}
  We explore the use of the Inverse Amplitude Method for unitarization
  of scattering amplitudes to derive the existence and properties of
  possible new heavy states associated with perturbative extensions of
  the electroweak breaking sector of the Standard Model starting from
  the low energy effective theory.  We use a toy effective theory
  generated by integrating out a heavy singlet scalar and compare the
  pole mass and width of the unitarized amplitudes with those of the
  original model.  Our results show that the Inverse Amplitude Method
  reproduces correctly the singlet mass up to factors of
  ${\cal O}$(1--3), but its width is overestimated.
\end{abstract}

\maketitle

\section{Introduction}
\label{sec:intro}

The discovery of a new particle~\cite{Aad:2012tfa,Chatrchyan:2012xdj}
resembling the Standard Model (SM) Higgs boson marks the beginning of
the direct study of the electroweak symmetry breaking sector
(EWSB). The complete characterization of the EWSB requires the precise
measurement of the Higgs couplings, as well as the search for new
states. In this work we analyze what we can learn from the observation
of departures from the SM predictions for the Higgs couplings in the
case that no new state is observed. As is well known, anomalous Higgs
couplings lead to rapid growth of the scattering amplitudes with
energy, leading to partial-wave unitarity
violation~\cite{Corbett:2014ora}.  Our goal is to verify how well
unitarization procedures, more specifically the Inverse Amplitude
Method (IAM)~\cite{Truong:1988zp,Dobado:1989qm,
  Dobado:1996ps,Oller:1997ng, GomezNicola:2001as} predict the
existence and properties of possible new states associated with
perturbative extensions of the SM. \smallskip

Here we consider the simplest extension of the SM symmetry breaking
system, {\it i.e.} we add a real singlet scalar field that is not
charged under the SM gauge group. Despite its simplicity, this
extension of the SM can have an impact in the Higgs physics at the
LHC~\cite{Barger:2007im,Bhattacharyya:2007pb,Bock:2010nz,Chen:2014ask},
as well as offering an interesting candidate for a portal to a
hidden sector~\cite{Binoth:1996au, Barbieri:2005ri, Patt:2006fw,
  Bowen:2007ia}. We assume that this singlet field is too heavy to be
produced so we integrate it out to obtain the low energy effective
theory.  \smallskip

The IAM is based on dispersion relations to unitarize the perturbative
partial wave amplitudes even in the presence of coupled channels, and
it has been applied with success to describe low energy hadronic
physics~\cite{Truong:1988zp,Dobado:1989qm, Dobado:1996ps,Oller:1997ng,
  GomezNicola:2001as}.  This method has also been extensively used to
study strongly interacting EWSB sectors and models exhibiting a heavy
Higgs~\cite{Dobado:1989gr, Delgado:2015kxa, Delgado:2014dxa,
  Espriu:2012ih, Espriu:2014jya}.  In this work we apply the IAM to
the effective theory generated by integrating out a heavy singlet
scalar and we compare its predictions to the original model
parameters.  In Sec.~\ref{sec:leff} we derive the corresponding
effective Lagrangian up to ${\cal O}(p^4)$ and, after briefly
reviewing the elements of the IAM relevant for our calculations in
Sec.~\ref{sec:iam}, we present our results and draw
  our conclusions in Sec.~\ref{sec:results}.  In particular we show
that the IAM indicates correctly that only the $I=0$ and $J=0$ channel
exhibits a resonance, reproducing the singlet mass up to factors of
${\cal O}$(1--3) even for relatively weak couplings. Its width,
however, is systematically overestimated.  \smallskip

\section{Effective Lagrangian for a Heavy Singlet Higgs Portal}
\label{sec:leff}

Our starting point is the SM scalar sector extended by a real singlet
scalar field $S$
\begin{equation}
{\cal L}(\Phi,S)=(D^\mu \Phi)^\dagger (D_\mu\Phi) +
\frac{1}{2}(\partial_\mu S)(\partial^\mu S) 
-V(\Phi,S)
\label{eq:Lsinglet0}
\end{equation}
where $\Phi$ stands for the SM scalar doublet and
\begin{equation}
V(\Phi,S) = - \mu^2_H \left|\Phi\right|^2 
+ \lambda \left|\Phi\right|^4 
- \frac{\mu^2_S}{2}\, 
S^2 + \frac{\lambda_S}{4}\, S^4 +  
\frac{\lambda_m}{2} \left|\Phi\right|^2 S^2\, .
\label{eq:Vsinglet1}
\end{equation}
For simplicity, we have imposed a $Z_2$ symmetry to forbid linear and cubic
terms in S. We concentrate on a scenario in which the S develops a
vacuum expectation value (vev) $v_S$. Presently, a new heavy scalar is
allowed provided the ratio of the SM vev ($v$) to $v_S$ is small and
the mixing between the mass eigenstates is small~\cite{Pruna:2013bma}.
\smallskip

As we will show below, in order to conveniently parametrize the low energy 
effective Lagrangian it is easier to write  the SM Higgs doublet as
\begin{equation}
\Phi = U(x)
\left ( 
\begin{array}{c}
0 \\
\frac{v+H}{\sqrt{2}}
\end{array}
\right )
\end{equation}
where $U$ is a function of the goldstone bosons 
$\omega_i(x)$
\begin{equation}
U(x)=\exp\left[\frac{i\omega(x)\cdot\tau}{v}\right]\;,
\end{equation}
and $\tau_i$ are the Pauli matrices. Therefore, we can write
Eq.~(\ref{eq:Lsinglet0}) as
\begin{eqnarray}
{\cal L}(H,S) &=& \frac{1}{2}(\partial_\mu H)(\partial^\mu H) 
+\frac{1}{2}(\partial_\mu S)(\partial^\mu S) +\frac{(v+H)^2}{4} 
{\rm Tr}[(D^\mu U) (D_\mu U)^\dagger] 
-\frac{1}{2} M_H^2 H^2 -\frac{1}{2} M_S^2 S^2 -\lambda_m v v_s 
H\, S\,
\nonumber
\\
&& 
-\left[
\lambda_S v_{s}\,S^3 + \frac{\lambda_S}{4} S^4\, 
+\frac{\lambda_m}{2} v_{s}\,  H^2 S  +  
\frac{\lambda_m}{4}(2vH+H^2)  S^2
+\lambda v H^3 
+\frac{\lambda}{4} H^4 
\right]
\label{eq:Lsinglet2}
\end{eqnarray}
with $M_H^2 = 2 \,\lambda \, v^2$, $ M^2_S = 2\, \lambda_S \, v_s^2$.
We have traded the  mass parameters $\mu^2_H$ and $\mu^2_S$ for the vev's
using the the minimization conditions,
$\mu^2_H = \lambda\, v^2 + \frac{\lambda_m}{2} v^2_{s} $, and
$\mu^2_S = \lambda_S v_{s}^2 + \frac{\lambda_m}{2} v^2$.  The
covariant derivative of $U$ takes the form:
\begin{equation}
D_\mu U(x)\equiv\partial_\mu 
U(x)
+\frac{i}{2}gW_\mu^a(x)\tau_a U(x)-\frac{ig^\prime}{2}B_\mu(x)
U(x)\tau_3\; . 
\label{eq:cdevsig}
\end{equation}

The two mass eigenstates $H_{1}$, and $S_1$ exhibit a doublet-singlet mixing
due to the presence of the $H\,S$ term in Eq.~(\ref{eq:Lsinglet2})
\begin{equation}
H_1=\cos\theta\, H\,+\sin\theta \,S  \;\; \;\; 
{\rm and}\;\; \;\;  S_1=\cos\theta\, S\,-\sin\theta \,H 
\end{equation}
with the lighter state ($H_1$) identified with the recently
discovered $125$ GeV Higgs particle. The mixing angle $\theta$ and
masses are given by \cite{Gorbahn:2015gxa}
\begin{equation}
\sin^2\theta = \frac{4\, y^2}{4\,
y^2 + \left(1 - x^2 + \sqrt{(1-x^2)^2 + 4 \,y^2}\right)^2}
\label{eq:mix}
\end{equation}
\begin{equation}
M^2_{H_1,S_1} = \frac{M_S^2}{2} \left(1 + x^2 \mp
\sqrt{(1-x^2)^2 + 4 \,y^2}\right)
\label{eq:mass} 
\end{equation}
with $x \equiv M_H/M_S $  and $y \equiv \lambda_m v/(2 \lambda_S v_S)$. 
\smallskip

In this scenario the heavier scalar $S_1$ is unstable and decays via
its mixing with the doublet or the singlet-doublet direct coupling in
Eq.~(\ref{eq:Lsinglet0}).  For $M_{S1}\geq 2 m_{\rm top}$ the $S_1$ width is
given by 
\begin{eqnarray}
\Gamma_{S_1}& = & 
\Gamma(S_1\rightarrow W^+ W^-)+
\Gamma(S_1\rightarrow Z Z)
+\Gamma(S_1\rightarrow t \bar t)
+\Gamma(S_1\rightarrow H_1 H_1) \nonumber \\
&=&
\frac{g^2 M_{S_1}^3}{128 \pi M_W^2}  \sin^2\theta \left[
2\left(1-x_W+\frac{3}{2} x_W^2\right)\sqrt{1-x_W}
+\left(1-x_Z+\frac{3}{2} x_Z^2\right)\sqrt{1-x_Z}
+3 x_t\sqrt{1-x_t}
\right]
\label{eq:widths1}
\\ \nonumber
&&+ \frac{\tilde{\lambda}^2}{32 \pi M_{S1}}  \sqrt{1-x_{H_1}}
 \; , 
\end{eqnarray} 
where $x_i=\frac{4 m_i^2}{M_{S_1}^2}$ and $\tilde{\lambda}/2$ is the
coefficient of the $S_1 H_1^2$ term obtained after we rotate
Eq.~(\ref{eq:Lsinglet2}) to the mass basis. Here, we are interested in
the scenario where $S_1$ is heavy compared with $H_1$ which allows us to
approximate Eq.~(\ref{eq:widths1}) by
\begin{equation}
\Gamma_{S_1} =
\frac{M_{S_1}}{256 \pi }  \Delta \left[
2\left(1-x_W+\frac{3}{2} x_W^2\right)\sqrt{1-x_W}
+\left(1-x_Z+\frac{3}{2} x_Z^2\right)\sqrt{1-x_Z}
+3 x_t\sqrt{1-x_t}
+ \sqrt{1-x_{H_1}}
\right]
 \; , 
\label{eq:widths2}
\end{equation}
where we have defined the parameter
\begin{equation}
\Delta =\frac{\lambda_m^2}{\lambda_S} \;\;{\rm so}
\;\;\;\Delta\, \frac{v^2}{2 M_{S_1}^2} \simeq \sin^2\theta\, .
\end{equation}

In the regime in which $S_1$ is very heavy we can integrate it out and
generate a low energy effective Lagrangian.  Since $H_1$ is not a
doublet field component, the corresponding effective Lagrangian cannot
be expressed in terms of higher-dimension operators obtained in the
linear representation of the electroweak symmetry breaking with a
doublet scalar.  As we will show below, it can, instead, be matched to
an effective chiral Lagrangian with a light Higgs\footnote{Alternatively 
if one integrates out the field $S$, ignoring
  the corrections due to mixing, one can match the resulting
  Lagrangian to an effective expansion in terms of higher-dimension
  operators involving the remaining doublet field $\Phi$ as is shown
  in the Appendix.}.\smallskip

We integrate out the $S_1$ field to obtain the tree-level effective
action using the  approach of Ref.~\cite{Henning:2014wua}: the
tree-level effective action is obtained by solving the equation of
motion (EOM) and inserting the solution into the action. In order to
do so we recast Eq.~(\ref{eq:Lsinglet2}) in the mass basis as
\begin{equation}
{\cal L}(H_1)+ \frac{1}{2} S_1 
\left[ -\partial^\mu \partial_\mu-M_{S_1}^2 -R(x)\right]S_1+ S_1 B(x) +
\Delta {\cal L}(H_1,S_1) 
\label{eq:LH1}
\end{equation}
with 
\begin{eqnarray}
B(x)= &&\frac{1}{4}{\rm Tr}[(D^\mu U) (D_\mu U)^\dagger] 
(H_1\sin2\theta+2 v \sin\theta)+
\frac{1}{4} H_1^3\left[\lambda_m \sin2\theta \cos 2\theta+2\sin 2\theta(\lambda_S \sin^2\theta-\lambda \cos^2\theta)\right] \nonumber \\
&& + \frac{1}{2} H_1^2\left[-3\sin 2\theta(\lambda v \cos\theta+\lambda_S v_S\sin\theta)+\lambda_m\sin 2\theta(v\cos\theta+v_S\sin\theta)-
\lambda_m(v\sin^3\theta+v_S\cos^3\theta)\right]
\nonumber\\
R(x)= && -\frac{1}{2}{\rm Tr}[(D^\mu U) (D_\mu U)^\dagger]\sin^2\theta
+\frac{1}{4} H_1^2\left[2\lambda_m  
\left ( \cos^2 2\theta -\frac{1}{2} \sin^22\theta \right)
+3\sin^2 2\theta
(\lambda+\lambda_S)\right] \nonumber \\
&&+H_1\left[\lambda_m(v\cos^3\theta -v_S \sin^3\theta)+
3\sin 2\theta(\lambda v \sin\theta-\lambda_S v_S \cos\theta)+
\sin 2\theta \lambda_m(v_S \cos\theta  -v \sin\theta )\right]\;.
\end{eqnarray}
$\Delta {\cal L}$ contains the non-quadratic terms $H_1 S_1^3$, $S_1^3$
and $S_1^4$. \smallskip

The linearized solution to the EOM for the field $S_1$ yields
\begin{equation}
S_{1C}=\frac{1}{\partial_\mu\partial^\mu +M_{S1}^2+R(x)} B(x)  \; .
\label{eq:sc10}
\end{equation}
Replacing $S_1$ by  $S_{1C}$ in Eq.~(\ref{eq:LH1}) one obtains
\begin{equation}
{\cal L}_{eff}(H_1)={\cal L}(H_1)+\frac{1}{2} B(x) S_{1C}+
\Delta {\cal L}(H_1,S_{1C}) \; .
\label{eq:eff0}
\end{equation}

Now we expand the effective Lagrangian (\ref{eq:eff0}) up to four
derivatives and keep only terms up to dimension-six which allows us to
match the resulting chiral effective Lagrangian to that of
Refs.~\cite{Alonso:2012px,Brivio:2013pma}\footnote{Notice
  despite that the UV theory is fully perturbative the effective low energy
  Lagrangian can be written as a theory more characteristic of strongly
  interacting/composite electroweak theories with a light scalar simply
  because these theories allow for enough freedom to account for the
  non-doublet nature of the light scalar.}
\begin{equation}
{\cal L}_{eff}(H_1)=
\frac{1}{2}(\partial_\mu H_1)(\partial^\mu H_1) 
-\frac{1}{2} M_{H_1}^2 H_1^2 + 
c_C\cP_C(H_1) + c_H\cP_H(H_1)+ c_6\cP_6(H_1) +c_7\cP_7(H_1)
- V(H_1)
\label{eq:lchiral}
\end{equation}
where 
\begin{equation}
\begin{aligned}
&\cP_C(H_1) 
= \frac{v^2}{4}
[\tr(D^\mu U) (D_\mu U)^\dagger] 
\cF_{C}(H_1)\,,
&&
\cP_H(H_1)
 = \frac{1}{2}(\partial^\mu H_1)(\partial_\mu H_1) \cF_H(H_1)\,,\\
&\cP_{6}(H_1) = [\tr(D^\mu U) (D_\mu U)^\dagger]^2 \cF_{6}(H_1) \,,
&&\cP_{7}(H_1) = [\tr(D^\mu U) (D_\mu U)^\dagger]  
\partial_\nu\partial^\nu\cF_{7}(H_1)  \,,
\end{aligned}
\label{eq:coefflchiral}
\end{equation}
and 
\begin{equation}
c_i \cF_i(H_1)\equiv c_i+ a_i \frac{H_1}{v} +b_i 
\left(\frac{H_1}{v}\right)^2+ d_i 
\left(\frac{H_1}{v}\right)^3 + e_i\left(\frac{H_1}{v}\right)^4 \; .
\label{eq:funcf}
\end{equation}

We present in Table~\ref{tab:coeff} the lowest non-zero order in
$v/M_{S_1}$ coefficients defining the functions $\cF$ in
Eq.~(\ref{eq:funcf}). Within our approximation, the $H_1$ potential is
given by
\begin{equation}
V(H_1)=\left(\lambda v-\frac{\lambda_m^2 v}{4\lambda_S}\right) H_1^3 
+\left(\frac{\lambda}{4}-\frac{\lambda_m^2}{16\lambda_S} \right) H_1^4
+{\cal O}\left(\frac{1}{M_{S_1}^2}\right) \;.
\end{equation}

\begin{table}
\begin{tabular}{l||c|c|c|c|c}
         & c & a & b & d & e\\\hline
$\cP_C(h)$ & 1 & 
$2-\frac{\lambda_m^2 v^2}{2 \lambda_S M_{S_1}^2}$ 
& $1-\frac{\lambda_m^2 v^2}{\lambda_S M_{S_1}^2}$   &
$-\frac{\lambda_m^2 v^2}{2 \lambda_S M_{S_1}^2}$   
& $\frac{\lambda_m^2\left
(9\lambda_m^2-16\lambda\lambda_S-10 \lambda_m\lambda_S\right)v^4}
{16 \lambda_S^2 M_{S_1}^4}$   \\ 
$\cP_H$ & 0 & 0 &   
$\frac{\lambda_m^2 v^2}{4 \lambda_S M_{S_1}^2}$   
& 0 & 0 \\
$\cP_6$ & $\frac{\lambda_m^2 v^4}{16 \lambda_S M_{S_1}^4}$   
 & $\frac{\lambda_m^2 v^4}{8 \lambda_S M_{S_1}^4}$   
& $\frac{\lambda_m^2 v^4}{16 \lambda_S M_{S_1}^4}$    & 0 & 0 \\   
$\cP_7$ & 0 & 0 & $\frac{\lambda_m^2 v^4}{8 \lambda_S M_{S_1}^4}$   
& 0 & 0  
\end{tabular}
\label{tab:coeff}
\caption{ Leading order in $v/M_{S_1}$
  coefficients defining the functions $\cF$ in
  Eq.~(\ref{eq:funcf}).}
\end{table}

It is interesting to notice that the operators generated at order
$p^4$ by the integration of $S_1$ modify the Higgs interactions with
electroweak gauge-boson pairs ($\cP_c$) and quartic electroweak
gauge-boson vertices ($\cP_6$), as well as introducing a rescaling
of all Higgs couplings to SM particles ($\cP_H$).

\section{Weak Gauge Boson Scattering and its Unitarization using the
  Inverse Amplitude Method}
\label{sec:iam}

The low energy effective Lagrangian in Eq.~(\ref{eq:lchiral}) implies
a modification of the gauge boson scattering with respect to the SM
expectation, leading to unitarity violation at high energies.  In this
respect, two of the operators generated are most relevant for this
discussion: $\cP_C(h)$ and $\cP_6(h)$.  $\cP_C(h)$ determines the
$H_1$ couplings to gauge boson pairs, in particular the term in $a_C$,
leads to a correction to the contribution of the virtual $H_1$
exchange required for unitarity. $\cP_6(h)$, in particular the term in
$c_6$, gives a contact four gauge boson coupling\footnote{$\cP_6(H)$
  without the Higgs terms, corresponds to the $L_5$ operator in
  Refs.\cite{Longhitano:1980tm,Appelquist:1994qz,Feruglio:1992wf} or
  ${\cal O}_5$ in Refs.\cite{Espriu:2012ih,Espriu:2014jya}, while
  $a_C$ and $b_C$ correspond respectively to the coefficients $2 a$
  and $b$ of, for example,
  Refs.~\cite{Contino:2010mh,Espriu:2014jya}.}.  \smallskip

For example, the scattering amplitude at tree level for longitudinal
gauge bosons is given by
\begin{equation}
 A \left( W^+_{L} W^-_{L} 
\rightarrow Z_{L} Z_{L} \right)= 
 A \left( W^+_{L} W^-_{L} 
\rightarrow Z_{L} Z_{L}\right)_{SM}+\left(
-\frac{1}{4}(a_C^2-4) \frac{v^2}{(s-M_{H_1}^2)}+8 a_6 \right)
\frac{(s-2M_W^2)(s-2M_Z^2)}{v^4} \:,
\label{eq:wwzzamp}
\end{equation}
where $\sqrt{s}$ is the center-of-mass energy. As we can see, the term
associated with $a_C$ grows as $s$ at high energy, while the one
containing $a_6$ exhibits growth with $s^2$, and hence lead to
violation of partial wave unitarity. \smallskip

The Inverse Amplitude Method (IAM)~\cite{Truong:1988zp} is an
approach, based on dispersion relations, that allows for the full
unitarization of the partial wave amplitudes. The IAM was originally developed for
chiral perturbation theory for mesons
~\cite{Dobado:1989qm,Dobado:1996ps,Oller:1997ng,GomezNicola:2001as}
and it was also applied to the unitarization of the one-loop weak
gauge boson scattering amplitudes without a light Higgs resonance
\cite{Dobado:1989gr}.  Most recently IAM has been applied in the
context of effective Lagrangians with a light Higgs
\cite{Espriu:2012ih,Espriu:2014jya,Delgado:2014dxa,Delgado:2015kxa},
mostly with the aim of inferring information about the possible
existence of heavier resonances associated with EWSB expected in
composite models with a new strongly interacting sector.  Let us
briefly summarize this approach. \smallskip

The rigorous derivation of the IAM is valid only for one or several
channels of particle pairs all with equal
masses~\cite{Delgado:2015kxa}.  In order to apply the IAM to the
longitudinal electroweak  gauge boson scattering one has to work in the isospin
symmetry approximation, {\it i.e.}  setting $c_{w} \to 1$
($M_Z \to M_W \equiv M$).  In this case if one defines the
longitudinally polarized weak-gauge boson scattering amplitudes as:
\begin{equation} 
A^{abcd}(p_1,p_2,p_3,p_4) \equiv 
A \left( W^{a}_{L}(p_1)W^{b}_{L}(p_2) 
\rightarrow W^{c}_{L}(p_3) W^{d}_{L}(p_4) \right) \, , 
\label{eq:basicamp}
\end{equation} 
one can construct the projected amplitudes ($T_I$)  with isospin $I$  as 
\begin{eqnarray}
T_{0} & = & \langle 00\vert S\vert 00 \rangle  =   
3 A^{+-00}(p_1,p_2,p_3,p_4) +   A^{++++}(p_1,p_2,p_3,p_4)  \;,
 \nonumber  \\
T_{1} & = & \langle 10\vert S\vert 10 \rangle  =   
2 A^{+-+-}(p_1,p_2,p_3,p_4) - 2 A^{+-00}(p_1,p_2,p_3,p_4) - 
A^{++++}(p_1,p_2,p_3,p_4) \;,
\label{eq:isosamp}
\\ \nonumber
T_{2} & = & \langle 20\vert S\vert 20 \rangle  =   
A^{++++}(p_1,p_2,p_3,p_4) \, . \nonumber
\end{eqnarray}
Isospin symmetry relates the amplitudes in  Eq.~(\ref{eq:isosamp})  
as
\begin{eqnarray} 
&&A^{+-+-}(p_1,p_2,p_3,p_4) 
=  A^{+-00}(p_1,p_2,p_3,p_4) 
+ A^{+-00}(p_1,-p_3,-p_2,p_4) \;,
\nonumber \\
&& A^{++++} (p_1,p_2,p_3,p_4) 
=A^{+-00}(p_1,-p_3,-p_2,p_4) 
+ A^{+-00}(p_1,-p_4,-p_3,p_2)   \,.
\end{eqnarray}
Defining $s=(p_1+p_2)^2$, $t=(p_1-p_3)^2=-\frac{1}{2}(s-4 M^2) (1-\cos\theta)$,
$u=(p_1-p_4)^2=-\frac{1}{2}(s-4 M^2)(1+\cos\theta)$ with $\theta$ the
scattering angle in the center of mass, we expand the isospin
amplitudes in partial waves as:
\begin{equation}
T_I=16 \pi \sum_j (2J+1) P_J(\cos\theta) t_{IJ}
\end{equation}
where the $P_{J}(x)$ are the Legendre polynomials. \smallskip

Let us assume that we know the isospin partial wave amplitudes
perturbatively as
\begin{equation}
t_{IJ}= t_{IJ}^{(0)} +t_{IJ}^{(2)}+ \dots \, ,
\label{eq:tijexp}
\end{equation}
where $t_{IJ}^{(0)}$ and $t_{IJ}^{(2)}$ are the leading order (LO) and
next to leading order (NLO) contributions.  
Then the IAM approximation~\cite{Truong:1988zp,Dobado:1989qm} 
of the full amplitude is
\begin{equation}
t_{IJ}\simeq 
t^{IAM}_{IJ} 
=\frac{t_{IJ}^{(0)}}{1-t_{IJ}^{(2)}/t_{IJ}^{(0)}}
=\frac{(t_{IJ}^{(0)})^2}{t_{IJ}^{(0)}-t_{IJ}^{(2)}}\; , 
\label{eq:iam}
\end{equation}
which, by construction, satisfies the unitarity constraint. \smallskip

In general, one has to deal with the possibility of coupled
channels~\cite{Oller:1997ng} . For instance, in the case of chiral
lagrangians applied to EWSB~\cite {Delgado:2014dxa,Delgado:2015kxa}
the processes $W_L^+ W_L^-\rightarrow h h$ and $hh\rightarrow hh$ also
contribute to the  partial wave $I=J=0$.  In this case one can
group the perturbatively expanded amplitudes in a matrix form as
\begin{equation}
M_{00}=M^{(0)}_{00}+M^{(2)}_{00} +\dots \equiv
\begin{pmatrix} 
t^{(0)}_{00}  &  t^{(0)}_{H,00}\cr
t^{(0)}_{H,0}  &  t^{(0)}_{HH,00}\cr
\end{pmatrix}
+
\begin{pmatrix} 
t^{(2)}_{00}  &  t^{(2)}_{H,00}\cr
t^{(2)}_{H,0}  &  t^{(2)}_{HH,00}\cr
\end{pmatrix}
+\dots \:,
\end{equation}
where we have introduced the isospin partial wave amplitudes $t_{H,00}$
and $t_{HH,00}$ corresponding to the processes 
$W_L^+ W_L^-\rightarrow h h$ and $hh\rightarrow hh$ respectively. In brief 
if we define

\begin{equation} 
A^{+-}(p_1,p_2,p_3,p_4)
= A(W^+_L(p_1)W^-_L(p_2)\to h(p_3)h(p_4)) \, ,
\end{equation}
whose projection in the $I=0$ channel is
\begin{equation}
\label{eq:fixed_isospin_hh}
T_{H, 0}= \sqrt{3} A^{+-}(p_1,p_2,p_3,p_4)\, 
\end{equation}
the relevant partial wave is
\begin{equation}
t_{H,00} = \frac{1}{64\pi} \int_{-1}^{1} d(\cos\theta)  T_{H,0} \, .
\end{equation}
And similarly $t_{HH,00}$ is the corresponding partial wave
amplitude for the $hh\rightarrow hh$ channel.  The unitarized
amplitude matrix in this case is~\cite{Delgado:2015kxa}
\begin{equation}
M^{IAM}_{00}=M^{(0)}_{00} 
\left( M^{(0)}_{00} -M^{(2)}_{00} \right)^{-1} M^{(0)}_{00} \: ,
\end{equation}
so the unitarized amplitude for the $WW\rightarrow WW$ channel is
\begin{equation}
t^{IAM}_{00} =\frac{ (t_{00}^{(0)})^2- t^{(0)}_{H,00}
\frac{t^{(0)}_{H,00} (t^{(0)}_{00}+t^{(2)}_{00})-2
t^{(2)}_{H,00} t^{(0)}_{00}}{t^{(0)}_{HH,00}-t^{(2)}_{H,00}}
}{t_{IJ}^{(0)}-t_{IJ}^{(2)}-\frac{(t^{(0)}_{H,00}-t^{(2)}_{H,00})^2
}{t^{(0)}_{HH,00}-t^{(2)}_{H,00}}}
\, ,
\label{eq:iamcoup}
\end{equation}
which clearly reduces to Eq.~(\ref{eq:iam}) if the amplitude of the mixed
channel ($t_{H,00}$) vanishes. \smallskip

Besides being a method for unitarization of the amplitudes, the
combination of terms appearing in the denominator of the IAM amplitude
allows for the possibility of having poles in the second Riemann sheet
for some regions of the parameter space. When they are close enough to
the physical region, those poles can be interpreted as {\sl
  resonances}. An alternative approach~\cite{Espriu:2014jya} to
identify these resonances appearing in the unitarized amplitudes is to
search for values of the center-of-mass energy ($\sqrt{s_{\rm pole}}$) for
which the real part of the denominator of the IAM amplitude $t^{IAM}_{IJ}$
vanishes, and then one identifies the mass of the resonance as
$M^2_R \equiv s_{\rm pole}$.  Expanding the amplitude near the pole as
\begin{equation}
t^{IAM}_{IJ} (s)\propto \frac{1}{(s-M_R^2)+i \sqrt{s}\Gamma_R} \;,
\label{eq:tpole}
\end{equation}
one can also derive the value of the resonance width as 
$\Gamma_R\propto {\rm Im} [t^{IAM}_{IJ} (s)]$.

\section{Results and Conclusions }\label{sec:results}

Next we apply the IAM to unitarize the gauge boson scattering
amplitudes obtained in the effective Lagrangian derived for the heavy
singlet Higgs portal model Eq.~(\ref{eq:lchiral}).  We will then
search for poles in the corresponding unitarized amplitudes and
reconstruct the properties of the inferred ``resonance(s)''.  In what
follows we will focus on the lowest $J$ partial wave amplitudes for
each isospin channel, {\it i.e.} $t_{00}$, $t_{11}$, and
$t_{20}$. \smallskip

Technically the mass and width of the ``reconstructed'' resonance are
obtained by searching for poles in the denominator of $t^{IAM}_{IJ}$ in
Eq.~(\ref{eq:iam}) {\it i.e.}  by solving
\begin{eqnarray}
t^{(0)}_{IJ}(M_R^2)-{\rm Re} \, t^{(2)}_{IJ}(M_R^2)
=0 \;\; \;\; \;\; \;\; \;\; 
&{\rm and}& \;\; \;\; \;\; \;\; \Gamma_R=-\frac{1}{M_R}\frac{{\rm Im}\,{t^{(2)}_{IJ}(M_R^2)}}
{\left.
\frac{d\left(t^{(0)}_{IJ}(s)-{\rm Re}\, t^{(2)}_{IJ}(s)\right)}{ds} 
\right|_{s=M_R^2}
}\;. 
\label{eq:masswidth}
\end{eqnarray}

In principle for $IJ=00$ we should consider the coupled channels,
which, as discussed in the previous section, are relevant to the
$WW \rightarrow WW$ scattering if $t_{H,00}$ is not too small.  For
large $s$, $t_{H,00}$ is proportional to $s [b_C-(a_C/2)^2]$
\cite{Delgado:2014dxa} and for the effective Lagrangian in
Eq.~(\ref{eq:lchiral}) this coefficient takes the value
$b_C-(a_C/2)^2=-\frac{\Delta}{{2}}
\frac{v^2}{M_{S_1}^2}$
which is assumed to be small in the effective Lagrangian expansion. So
the inclusion of the coupled channels represents a small correction
which, for simplicity, we neglect in the following and we search for
the resonances in the $IJ=00$ channel as in 
Eq.~(\ref{eq:masswidth})\footnote{We have also verified that if we
  artificially set $b_C=a_C^2$ in our calculations the reconstructed
  value of mass and width found in the $IJ=00$ channels are very similar
  to those obtained with the correct value.}.  The effect of the
$WW\rightarrow hh$ channel is, nevertheless, taken into account in the
evaluation of ${\rm Im}\, t^{(2)}_{00}$ (see Eq.~(\ref{eq:optical})
below). \smallskip

In this work, we evaluate tree level amplitudes using FeynArts
\cite{Hahn:2000kx} with the anomalous Higgs interactions from the
Lagrangian Eq.~(\ref{eq:lchiral}) introduced using
FeynRules~\cite{Christensen:2008py} and take the exact isospin
limit.  Our results agree with the expressions in the literature
\cite{Espriu:2012ih} in the corresponding limits. \smallskip


In order to organize the perturbative expansion of the $t_{IJ}$ we
follow the counting in terms of powers of $p$ that is characteristic
of chiral Lagrangians~\cite{Buchalla:2013rka}.  In this expansion the
tree-level contributions from the Higgs anomalous couplings, $a_C-2$,
and $b_C-1$, are counted as being part of $t^{(0)}_{IJ}$, {\it i.e.}
${\cal O}(p^2)$, and therefore their corresponding loop contributions
must be included in $t^{(2)}_{IJ}$ since they are ${\cal O}(p^4)$.  To
include the loop corrections in
${\rm Re}\, t^{(2)}_{IJ}$ we follow the approach of
Ref.~\cite{Espriu:2014jya} and use the expressions obtained in
Ref.~\cite{Espriu:2013fia} using the equivalence
theorem~\cite{Cornwall:1974km,Lee:1977eg} and given in the
approximation of massless external particles.  The divergent part of
these loops cancel against the renormalization of some of the
tree-level couplings of the ${\cal O}(p^4)$ operators defined at some
renormalization scale $\mu_R$.  This is the case of $c_6$ which then
at a scale $s$ becomes
\begin{equation}
c_6(s)\simeq c_6(\mu_R^2)-\frac{1}{24}\frac{1}{4\pi}\left[
\left(1-\frac{a^2_C}{4}\right)^2+\frac{3}{2}\left(
\left(1-\frac{a^2_C}{4}\right)^2-(1-b_C)^2\right)^2\right]
\log\frac{s}{\mu_R^2} \;.
\end{equation}
In our calculations we will take the renormalization scale as the mass
of the heavy scalar $\mu_R=M_{S_1}$. Thus when extrapolating the
amplitudes to scales $s\sim M^2_{S_1}$ we can approximate
$c_6(s)\simeq c_6(M_{S_1}^2)$ with $c_6(M^2_{S_1})$ given in
Eq.~(\ref{eq:coefflchiral}) and in Table~I\footnote{The same loops
  generate a coefficient for the operator
  $[\tr(D^\mu U) (D^\nu U)^\dagger][\tr(D_\mu U) (D_\nu U)^\dagger]$.
  Such an operator is not generated by integrating out $S_1$ at the order
  given in Eq.~(\ref{eq:lchiral}). Thus we will take the corresponding
  renormalized coefficient to be zero in our calculations.}.  The
remaining finite part of the loop from both the SM and the anomalous
values of $a_C$ and $b_C$ is included in $t^{(2)}$ as well. In
order to estimate the uncertainty associated with the approximations
used in the evaluation of this finite part we have performed our
calculations both including and excluding the anomalous loop
contribution. We will refer to these two calculations as
${\cal O}(p^4)$-1loop and ${\cal O}(p^4)$-tree
respectively. \smallskip

We first look for the presence of physical poles in the isospin
amplitudes $t^{IAM}_{IJ}(s)$ as a function of the relevant parameters
of the effective Lagrangian: the coupling ratio $\Delta$ and the mass
scale $M_{S_1}$ which determine the values of all relevant anomalous
couplings entering the $WW \rightarrow WW$ scattering, in particular
$a_C$, $b_C$ and $c_6$.  One must notice that for the simplified
potential in Eq.~(\ref{eq:Vsinglet1}) the condition that the
electroweak breaking minimum is a global minimum sets an upper bound
for $\Delta < 4\lambda\simeq 0.6$; see Ref.~\cite{Chen:2014ask} for a
recent analysis of the bounds with a more general
potential. Nevertheless, in what follows, we will extend our study to
larger values of $\Delta$ to illustrate the results in stronger
coupled scenarios. \smallskip

\begin{figure}
\includegraphics[width=0.9\textwidth]{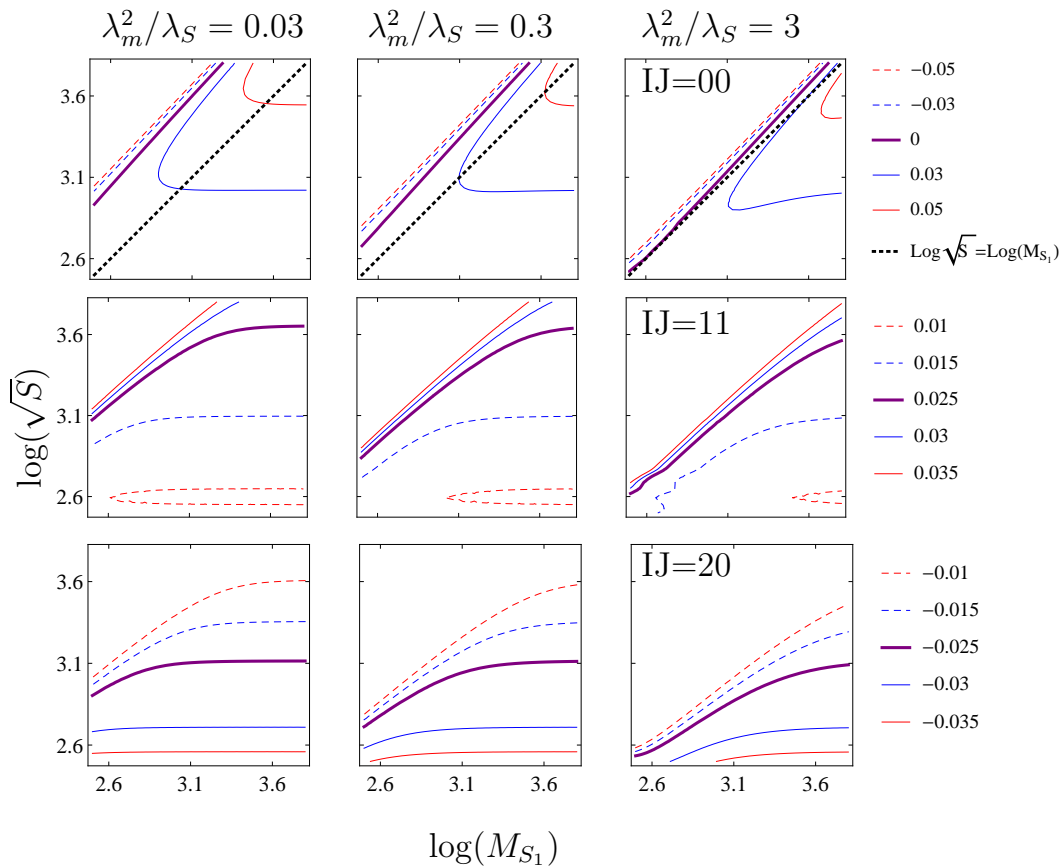}
\caption{Contours of the functions
  ${\rm Re}[t^{(0)}_{IJ}(s)-t^{(2)}_{IJ}(s)]$ in the plane
  $(\sqrt{S},M_{S_1})$ for three characteristic values of the relevant
  coupling ratio $\Delta=\lambda_m^2/\lambda_S$ and for the three
  isospin channels $IJ=00$ (upper panels), $IJ=11$ central panels, and
  $IJ=20$ (lower panels).}
\label{fig:contours}
\end{figure}

We show in Fig.~\ref{fig:contours} contours of the real part of the
denominator of the ${\cal O}(p^4)$-1loop functions $t^{IAM}_{IJ}(s)$,
{\it i.e.}  ${\rm Re}(t^{(0)}_{IJ}-t^{(2)}_{IJ})$, for $IJ=00$ (upper
panels), 11 (central panels), and 20 (lower panels) in the
$s \otimes M_{S_1}$ plane and for three characteristic values of
$\Delta=0.03$, $0.3$, and $3$.  Therefore, this figure illustrates
that for no value of $\Delta$ do the functions
${\rm Re}(t^{(0)}_{11}-t^{(2)}_{11})$ and
${\rm Re}(t^{(0)}_{20}-t^{(2)}_{20})$ present a zero in the physical
plane, while ${\rm Re}\,(t^{(0)}_{00}-t^{(2)}_{00})$ as a function of
$s$ always possesses a zero for any value of $\Delta$ and $M_{S1}$.
In other words, the effective theory after unitarization by the IAM
method is compatible with the presence of one possible physical scalar
resonance in the zero-isospin channel and none in any other
spin-isospin channels, which is in agreement with the original full
theory that has a scalar $S_1$ state in the physical spectrum and no
other heavy states. \smallskip

For the sake of illustration, we also present in the upper panels of
Fig.~\ref{fig:contours} the line corresponding to $\sqrt{s}=M_{S_1}$
for comparison with the zero value contour which determines the
position of the resonance $s=M_{R}$. As seen in this figure, the
larger the value of $\Delta$ the closer the two lines, {\it i.e.}  the
reconstructed mass of the IAM resonance is closer to the real mass of
the scalar of the full theory for stronger couplings. The results in
the figure correspond to the ${\cal O}(p^4)$-1loop calculation, but
the same qualitative results hold for the ${\cal O}(p^4)$-tree
calculation. \smallskip

\begin{figure}
\includegraphics[width=0.7\textwidth]{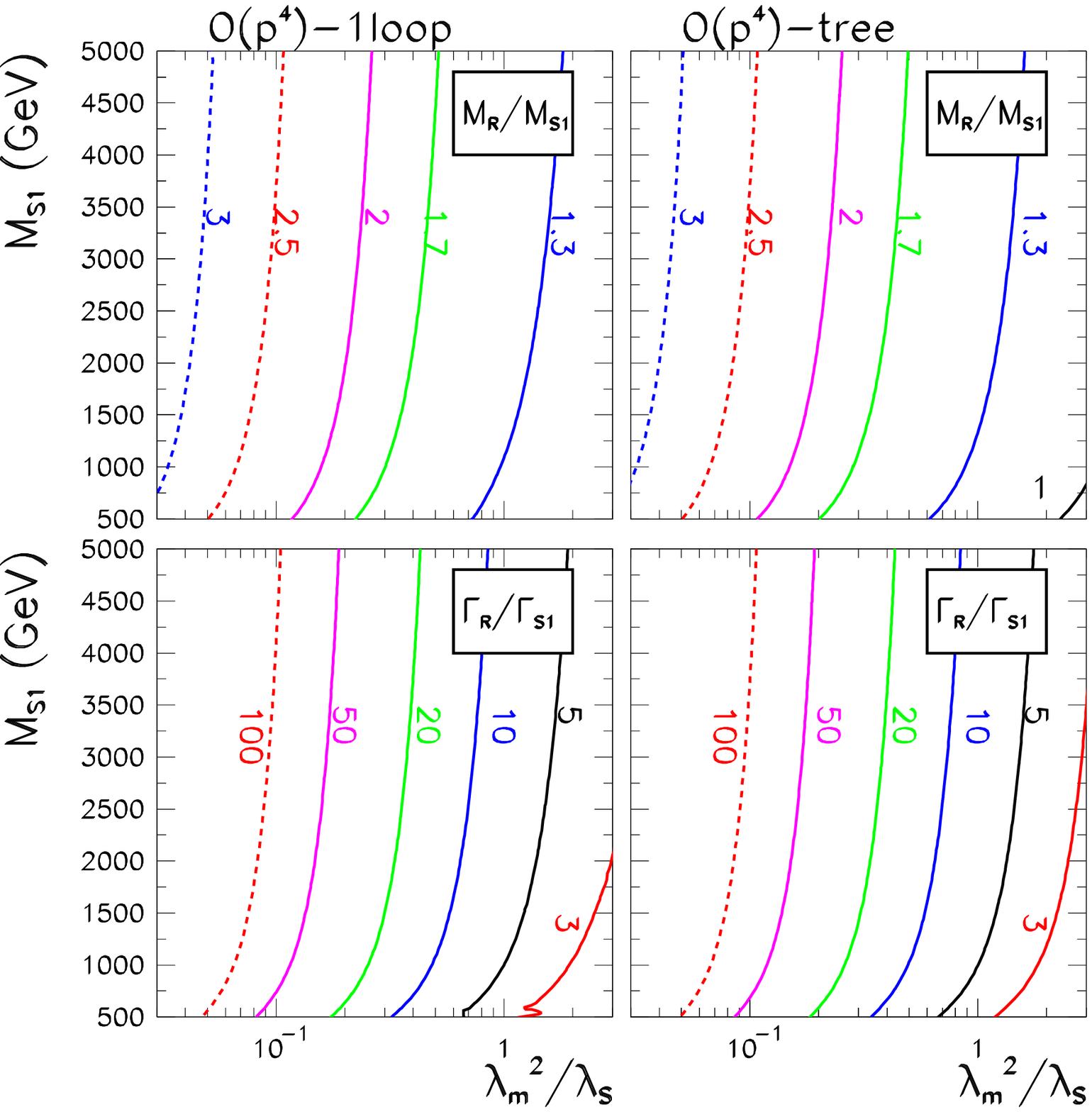}
\caption{{\bf Upper panels:} Contours of the ratio of the mass of the
  resonance found in the $t_{00}$ channel, $M_R$, to the mass of the
  integrated out scalar, $M_{S_1}$, versus the relevant ratio of
  Yukawa couplings $\Delta=\lambda_m^2/\lambda_S$ and
    $M_{S_1}$.  {\bf Lower panels:} Contours the ratio of the width of
  the resonance found in the $t_{00}$ channel, $\Gamma_R$, to the
  width of the scalar, $\Gamma_{S_1}$ in the plane
  $\Delta \otimes M_{S1}$.  }
\label{fig:ratios}
\end{figure}

We further quantify this comparison in Fig.~\ref{fig:ratios} where
the upper panels depict the ratio of the reconstructed scalar pole
mass $M_R$ over the $S_1$ mass as a function of $\Delta$
and $M_{S_1}$ for the ${\cal O}(p^4)$-1loop
calculation (left upper panel) and ${\cal O}(p^4)$-tree calculation
(right upper panel). As seen in these panels, the masses agree
within a factor ${\cal O} (1-3)$, even for very small couplings
independent of whether the approximate one-loop or tree amplitudes
are included in the calculation.\smallskip

In order to verify that the scalar pole found can be interpreted as a
physical state we also compute its width as in
Eq.~(\ref{eq:masswidth}).  For this we need to evaluate
${\rm Im}\, t^{(2)}_{00}(s)$ which could be obtained from
the 1-loop result in Ref.~\cite{Espriu:2013fia}
  obtained using the equivalence theorem.
However, since the 1-loop amplitude is only known approximately, we
estimate the relevant imaginary part by perturbative application of
the optical theorem
\begin{equation}
{\rm Im \,} 
t^{(2)}_{00}(s) =
\frac{2 p}{\sqrt{s}}  |t^{(0)}_{00}(s)|^{2}+
\frac{2 p_H}{\sqrt{s}}  |t^{(0)}_{H,00}(s)|^{2} \; , 
\label{eq:optical}
\end{equation}
where $p$ ($p_H$) is the modulus of the three-momentum of the gauge
bosons ($H_1$ pairs) in the center of mass.

We find that for all values of the model parameters $\Delta$ and
$M_{S_1}$ the reconstructed width is positive, so the interpretation
of the amplitude pole as a physical scalar state with a mass
relatively close to the real scalar mass $M_{S_1}$
holds. Notwithstanding, when compared with the perturbatively computed
$S_1$ width in Eq.~(\ref{eq:widths2}) we find that the reconstructed
width is considerably larger as seen in the lower panels in
Fig.~\ref{fig:ratios}, particularly for the more weakly interacting
scenarios. This is somehow not unexpected. The IAM method was built to
unitarize strong interaction amplitudes for which the
resonance-dominance approximation holds and the amplitude near the
pole of a resonance is fully determined by the resonance mass and
width. However, for weakly interacting scenarios, such as that
used here for illustration, the violation of unitarity is relatively
mild and the full amplitude, even near the new state, contains a
non-negligible ``continuous'' contribution from the SM.  So the
factorization used in Eqs.~(\ref{eq:tijexp}) and (\ref{eq:iam}) with
the full SM contribution included in the reconstructed amplitude as
part of the resonance amplitude does not seem to be optimum.
\smallskip

{\sl In summary}, in this work we have explored the capability of the
Inverse Amplitude Method for unitarization of scattering amplitudes to
predict the properties of possible new heavy states associated with
perturbative electroweak breaking extensions of the SM, using as
a starting point the unitarity violating amplitudes of the low energy
effective theory.  We have used as a study case that of the singlet
Higgs portal. First in Sec.~\ref{sec:leff} we derived the
effective Lagrangian obtained after integrating out the heavier scalar
while leaving the lighter scalar, a mixture of the doublet and
singlet states. We showed that in this case the effective Lagrangian can
be matched to that of a chiral expansion which we write up to
${\cal O}(p^4)$. With this effective Lagrangian in hand we 
obtained the relevant unitarity violating amplitudes.  Working in the
isospin approximation, we used the IAM method to reconstruct
unitarized amplitudes and search for possible physical poles 
in these amplitudes. The results in Sec.~\ref{sec:results} show that only the
unitarized spin scalar zero-isospin amplitude presents poles in the
physical plane, in agreement with the full theory which has only one
additional heavy scalar. We also find that the IAM reconstructs
correctly the scalar singlet mass up to factors of ${\cal O}$(1--3)
even for relatively weak couplings. Nevertheless its width is
systematically overestimated.

\section*{Acknowledgments}
We thank J. Taron for careful reading of the manuscript and discussions.
M.C. G-G and O.J.P.E are greatful to the CERN theory group for their
generous hospitality during part of the develepment of this work.
O.J.P.E. is supported in part by Conselho Nacional de Desenvolvimento
Cient\'{\i}fico e Tecnol\'ogico (CNPq) and by Funda\c{c}\~ao de Amparo
\`a Pesquisa do Estado de S\~ao Paulo (FAPESP); M.C.G-G and T.C. are
supported by USA-NSF grant PHY-13-16617 and by FP7 ITN INVISIBLES
(Marie Curie Actions PITN-GA-2011-289442).  M.C.G-G also acknowledges
support by grants 2014-SGR-104 and by FPA2013-46570 and
consolider-ingenio 2010 program CSD-2008-0037. T.C. was supported in part
by the Australian Research Council.


\appendix

\section{Effective Lagrangian after Integrating out the $S$ Field}

Our starting point is the SM extended by the addition of a real
singlet scalar field $S$ as given in Eqs.~(\ref{eq:Lsinglet0}) and
(\ref{eq:Vsinglet1}).  Below the scale at which $S$ acquires a vev
$v_s$ the scalar potential can be written as~\cite{Gorbahn:2015gxa}
\begin{equation}
\label{eq:Vsinglet02}
V(\Phi,S) = - \tilde \mu^2_H\left|\Phi\right|^2 
+ \lambda \left|\Phi\right|^4 + 
\frac{M_S^2}{2} S^2 
+ v_{s}\lambda_S \,S^3 + \frac{\lambda_S}{4}\, 
S^4 +\lambda_m v_{s}\, \left|\Phi\right|^2 S 
+  \frac{\lambda_m}{2} \left|\Phi\right|^2 S^2\, ,
\end{equation}
with $\tilde \mu^2_H = \mu^2_H - (\lambda_m v_{s}^2) /2$ and
$M_S^2=2 \lambda_S v^2_S$ .  Now we rewrite this Lagrangian as
$\mathcal{L} = \mathcal{L}(\Phi) + \Delta\mathcal{L}(\Phi,S) $ with
\begin{equation}
\Delta \mathcal{L} = \frac{1}{2} \left(\partial_\mu
S\right)^2 - \frac{1}{2} M_S^2 S^2 -A \left| \Phi \right|^2 S -
\frac{1}{2} k\left| \Phi \right|^2 S^2 - \frac{1}{3!}\mu S^3
-\frac{1}{4!}\tilde \lambda_S S^4, 
\label{eq:Lsinglet}
\end{equation} 
where
\begin{equation}
\mu=6\, \lambda_S v_S\;,\;\;
\tilde\lambda_S=6\, \lambda_S \;, \;\;
k=\lambda_m\;,\;\;
A=\lambda_m v_S\;. 
\label{eq:notrip}
\end{equation}

In order to apply the tree level integration procedure described in
Sec.~\ref{sec:leff} for the singlet field $S$ we must solve the EOM
for the $S$ field at lowest order leading to
\begin{equation}
S_C=\frac{1}{\partial_\mu\partial^\mu +M_S^2+U} B 
\label{eq:sc0}
\end{equation}
where we have defined
\begin{equation}
B=-A \, |\Phi^2|
 \;\;\;\; U= k \, |\Phi^2| \,.
\end{equation}
Introducing $S_C$ in Eq.~(\ref{eq:Lsinglet}) and keeping the terms up
to order dimension-eight, one obtains the following anomalous interactions
\begin{eqnarray}
\Delta \mathcal{L}_{\rm eff}&=&
\frac{A^2}{2M_S^2} |\Phi|^4
+\frac{A^2}{2M_S^4}
\partial_\mu |\Phi|^2\partial^\mu |\Phi|^2
+\frac{A^2}{2M_S^4}\left(\frac{A\mu}{3M_S^2}-k\right) |\Phi|^6\\ \nonumber
&+&\frac{A^2}{2M_S^6}
\left(-\frac{A^2 \tilde\lambda_S}{12M_S^2}+k^2
-\frac{A\,\mu\, k}{M_S^2} \right)|\Phi|^8 +
\frac{2 A^2}{M_S^6}\left(\frac{A\mu}{2 M_S^2} -k\right)
|\Phi^2|\partial_\mu |\Phi|^2\partial^\mu |\Phi|^2
+ \frac{A^2}{2M_S^6}
\partial_\mu\partial^\mu |\Phi|^2
\partial_\nu\partial^\nu|\Phi|^2\; . 
\end{eqnarray}

At this point it is interesting to apply the EOM of the doublet field
to the last term of the equation above to better observe the emergence
of an anomalous quartic coupling between the electroweak gauge bosons.
This is possible since the invariance of the physical observables
under the associated operator redefinitions is guaranteed as it has
been proven that operators connected by the EOM lead to the same
$S$--matrix elements~\cite{equiv-s}.  The EOM for the doublet field
reads
\begin{equation}
(D^\mu D_\mu \Phi)=  \mu_H^2 \Phi - 2 \lambda \Phi |\Phi|^2+  
F_{\rm ferm} \,
\end{equation}
where $F_{\rm ferm}$ is a function involving  fermionic fields
from the Yukawa operators. Moreover, using this EOM 
we find
\begin{equation} 
\partial_\mu\partial^\mu |\Phi|^2
=2 \left[(D^\mu\Phi)^\dagger (D_\mu \Phi)
+  \mu_H^2 |\Phi|^2 
-2\lambda |\Phi|^4\right]+  \hbox{ terms containing fermionic fields}
\end{equation}
Therefore, altogether we find that for terms involving only scalar
and/or gauge bosons
\begin{eqnarray}
\Delta \mathcal{L}_{\rm eff}&=&
 -\Delta\lambda |\Phi|^4+\frac{f_{\Phi,3}}{M_S^2} {\cal O}_{\Phi,3}
+\frac{f_{\Phi,5}}{M_S^4} {\cal O}_{\Phi,5}
+\frac{f_{\Phi,2}}{M_S^2} {\cal O}_{\Phi,2}
+\frac{f_{\Phi,4}}{M_S^2} {\cal O}_{\Phi,4}
+\frac{f_{\Phi,6}}{M_S^4} {\cal O}_{\Phi,6}
+\frac{f_{\Phi,7}}{M_S^4} {\cal O}_{\Phi,7}
+\frac{f_{S,1}}{M_S^4} {\cal O}_{S,1}
\label{eq:dim8lag}
\end{eqnarray}
with $\Delta \lambda=-\frac{A^2}{2 M_S^2} \left(
  1+\frac{4\mu_H^4}{M_S^4}\right)=
-\frac{\lambda_m^2}{4\lambda_S}
  \left( 1+\frac{4\mu_H^4}{M_S^4}\right)$ and
\begin{equation}
\begin{array}{lll}
{\cal O}_{\Phi,2}=
\frac{1}{2}\partial_\mu |\Phi|^2\partial^\mu |\Phi|^2 &&
f_{\Phi,2}=\frac{A^2}{M_S^2} = \frac{\lambda_m^2}{2\lambda_S} 
\nonumber \\
{\cal O}_{\Phi,3}=\frac{1}{3}|\Phi|^6 &&
f_{\Phi,3}=
\frac{3A^2}{2M_S^2}\left(\frac{A\mu}{3M_S^2}-k
-\frac{16 \lambda \mu_H^2}{M_S^2}\right)=  
-\frac{12 \lambda_m^2}{\lambda_S}
\frac{\lambda\mu_H^2}{M_S^2}
\nonumber \\
{\cal O}_{\Phi,4}=
(D^\mu\Phi)^\dagger (D_\mu \Phi) |\Phi|^2 &&
f_{\Phi,4}=\frac{4A^2\mu_H^2}{M_S^4} =
\frac{2\lambda_m^2}{\lambda_S}\frac{\lambda\mu_H^2}{M_S^2} 
\nonumber \\
{\cal O}_{\Phi,5}
=\frac{1}{4}|\Phi|^8 && 
f_{\Phi,5}= 
\frac{2A^2}{M_S^2}
\left(-\frac{A^2 \tilde\lambda_S}{12M_S^2}+k^2
-\frac{A\, k\,\mu}{M_S^2} +16 \lambda^2\right)  
\simeq (64 \lambda^2-9\lambda_m^2)\frac{\lambda_m^2}{4\lambda_S}
\nonumber \\
{\cal O}_{\Phi,6}=
\frac{1}{2} |\Phi|^2\partial_\mu |\Phi|^2\partial^\mu |\Phi|^2 &&
f_{\Phi,6}=\frac{4 A^2}{M_S^2}\left(\frac{A\mu}{2 M_S^2} -k\right)
=\frac{\lambda_m^3}{\lambda_S}
\nonumber \\
{\cal O}_{\Phi,7}=|\Phi|^2
(D^\mu\Phi)^\dagger (D_\mu \Phi) |\Phi|^2 &&
f_{\Phi,7}=-\frac{8 A^2 \lambda}{M_S^2}=-4 \lambda\frac{\lambda_m^2}{\lambda_S}
 \nonumber \\
{\cal O}_{S,1}= (D^\mu\Phi)^\dagger (D_\mu \Phi) 
(D^\nu\Phi)^\dagger (D_\nu \Phi) &&
f_{S,1}= \frac{2 A^2}{M_S^2}=\frac{\lambda_m^2}{\lambda_S}
\end{array}
\end{equation}
Notice that in the last column we have introduced the relations in
Eq.~(\ref{eq:notrip}) and we have expanded to the lowest non-zero
order in $\mu_H^2/M_S^2$. \smallskip

The effective Lagrangian in Eq.~(\ref{eq:dim8lag}) also leads to
violation of unitarity of electroweak boson scattering. For example
the $W^+W^-\rightarrow Z Z$ amplitude takes the form in
Eq.~(\ref{eq:wwzzamp}) with the identification (at the lowest order in
inverse powers of the heavy mass) $a_C=2-f_{\Phi,2} \frac{v^2}{M_S^2}$
and $c_6=\frac{f_{S,1}}{16} \frac{v^4}{M_S^4}$~\cite{Eboli:2006wa} .

\end{document}